\documentclass[iop,apj]{emulateapj}
\usepackage{hyperref}
\hypersetup{breaklinks,
colorlinks,
linkcolor=blue,
citecolor=blue,
urlcolor=blue}
\usepackage{journals}

\begin{document}

\shorttitle{Laboratory WD Photospheres}
\shortauthors{Falcon et al.}

\title{Laboratory Measurements of White Dwarf Photospheric Spectral Lines: H$\beta$}

\author{Ross~E.~Falcon\textcolor{blue}{\footnotemark[1,2,*]}}
\author{G.~A.~Rochau\textcolor{blue}{\footnotemark[2]}}
\author{J.~E.~Bailey\textcolor{blue}{\footnotemark[2]}}
\author{T.~A.~Gomez\textcolor{blue}{\footnotemark[1,2]}}
\author{M.~H.~Montgomery\textcolor{blue}{\footnotemark[1]}}
\author{D.~E.~Winget\textcolor{blue}{\footnotemark[1]}}
\author{T.~Nagayama\textcolor{blue}{\footnotemark[2]}}

\affil{\footnotemark[1]Department of Astronomy and McDonald Observatory, University of Texas, Austin, TX, 78712
\vspace{0.5mm}\\
\footnotemark[2]Sandia National Laboratories, Albuquerque, NM 87185-1196, USA}

\footnotetext[*]{\href{mailto:refalco@sandia.gov}{refalco@sandia.gov}}

\begin{abstract}
We spectroscopically measure multiple hydrogen Balmer line profiles from laboratory plasmas to investigate the theoretical line profiles used in white dwarf atmosphere models.  X-ray radiation produced at the {\it Z} Pulsed Power Facility at Sandia National Laboratories initiates plasma formation in a hydrogen-filled gas cell, replicating white dwarf photospheric conditions.  Here we present time-resolved measurements of H$\beta$ and fit this line using different theoretical line profiles to diagnose electron density, $n_{\rm e}$, and $n=2$ level population, $n_2$.  Aided by synthetic tests, we characterize the validity of our diagnostic method for this experimental platform.  During a single experiment, we infer a continuous range of electron densities increasing from $n_{\rm e}\sim4$ to $\sim30\times10^{16}\,$cm$^{-3}$ throughout a 120-ns evolution of our plasma.  Also, we observe $n_2$ to be initially elevated with respect to local thermodynamic equilibrium (LTE); it then equilibrates within $\sim55\,$ns to become consistent with LTE.  This supports our electron-temperature determination of $T_{\rm e}\sim1.3\,$eV ($\sim15,000\,$K) after this time.  At $n_{\rm e}\gtrsim10^{17}\,$cm$^{-3}$, we find that computer-simulation-based line-profile calculations provide better fits (lower reduced $\chi^2$) than the line profiles currently used in the white dwarf astronomy community.  The inferred conditions, however, are in good quantitative agreement.  This work establishes an experimental foundation for the future investigation of relative shapes and strengths between different hydrogen Balmer lines.
\end{abstract}


\section{Introduction}\label{intro}

Recently, \citet[][hereafter TB]{Tremblay09} revamped white dwarf (WD) atmosphere models by modifying the constituent theoretical hydrogen line profiles, which are calculated using the Unified Theory of Stark broadening \citep[][hereafter VCS]{Smith69,Vidal70,Vidal71,Vidal73}.  They followed the lead of \citet{Seaton90} to incorporate the occupation probability formalism of \citet{Hummer88} into the line-profile calculation to account for the loss of upper-energy-level transitions due to the high electric microfields common in WD-atmosphere plasmas.  This significantly improves the consistency of effective temperature, $T_{\rm eff}$, and surface gravity, log\,$g$, inferred from different, spectroscopically observed hydrogen Balmer lines.  It also results in a systematic increase in both $T_{\rm eff}$ and log\,$g$.

TB used their line profiles to spectroscopically determine an increased mean mass of $\langle M\rangle=0.649\,M_\odot$ for DA WDs from the Palomar-Green Survey \citep{Liebert05}.  \citet{Falcon10} used gravitational redshifts -- a method mostly independent of theoretical line profiles -- to infer a DA mean mass of $\langle M\rangle=0.647^{+0.013}_{-0.014}\,M_\odot$ from the European Southern Observatory supernovae Ia progenitor survey \citep{Napiwotzki01an}.  This is in excellent agreement with TB.  Subsequent spectroscopic studies using TB line profiles, however, do not find mean masses from other WD samples that agree with that determined from \citet{Falcon10}.  These include \citet{Limoges10}, \citet{Tremblay11}, and \citet{Kleinman13}, who found $\langle M\rangle=0.606$, $0.613$, and $0.623\,M_\odot$, respectively.

The impact of theoretical line profiles on the interpretation of observed WD spectra as well as the discrepancy between the inferred mean masses from the spectroscopic and gravitational-redshift methods motivate the development of laboratory experiments capable of testing line-profile theories.  TB profiles now replace the tabulated VCS profiles of \citet{Lemke97} as the WD-astronomy standard.  The difference between these calculations increases with increasing principal quantum number, $n$, and with electron density, $n_{\rm e}$.  Therefore, measurements of the {\it relative} line {\it shapes} of hydrogen Balmer lines formed at high electron densities can be used to discriminate between theoretical line-profile models.  However, few laboratory experiments push into high $n_{\rm e}$ ($\gtrsim10^{17}\,$cm$^{-3}$) for the investigation of H line shapes \citep[e.g.,][]{Morris68,Parigger03,Parigger08,Djurovic05,Djurovic09}, and none of these measure multiple lines in the same plasma.  Simultaneous measurements of multiple lines is highly desirable because it alleviates potential systematic uncertainties and promotes measurement accuracy.

Our experiments create plasmas at WD photospheric conditions and reaches $n_{\rm e}>10^{17}\,$cm$^{-3}$ while simultaneously observing multiple spectral lines \citep{Falcon10b,Falcon13b,Falcon13,Montgomery15}.  Hence, we provide the first such laboratory data at these plasma conditions; we do so using a fundamentally different experimental approach than previous studies \citep[e.g.,][]{Wiese63,Hill67,Bengtson69,Helbig81} by radiatively driving plasma formation \citep[e.g.,][]{Mancini09,Liedahl11}.  Because we use time-resolved spectroscopy, we also measure throughout a continuous range of $n_{\rm e}$ in a single experiment and from the same plasma.

Furthermore, we observe our plasma in absorption.  This, along with simultaneously observing multiple lines, provides a unique opportunity to experimentally explore occupation probabilities \citep{Hummer88} by measuring relative line {\it strengths}.

In this paper we present spectroscopic measurements of the H$\beta$ line.  Historically, theoretical calculations, particularly widths, of this line agree with measurements from benchmark laboratory experiments \citep[e.g.,][]{Wiese72}, validating its accuracy (for $n_{\rm e}<10^{17}\,$cm$^{-3}$) to diagnose plasma conditions \citep[e.g.,][]{Kelleher93}.  We thus fit H$\beta$ to infer electron density, $n_{\rm e}$, and $n=2$ level population, $n_2$, finding agreement between different theoretical line-profile models (Section~\ref{fits}).  These $n_{\rm e}$ and $n_2$ measurements also allow us to infer electron temperature, $T_{\rm e}$ (Section~\ref{measure_electron_temperature}).  We then scrutinize our sensitivity to experimental uncertainties.  Section~\ref{electron_temperature} synthetically investigates the dependence on $T_{\rm e}$, and Sections~\ref{transverse_gradient} and \ref{LOS_gradient} investigate inhomogeneities or gradients in the plasma conditions.  Our scope is to secure our measured H$\beta$ line as a diagnostic anchor for the investigation of relative line profiles ({\it shapes} and {\it strengths}) as part of the laboratory study of WD photospheres.

\section{Experimental Platform}\label{platform}

Ours is part of the {\it Z} Astrophysical Plasma Properties (ZAPP) Collaboration \citep{Rochau14}, a group of experiments conducted simultaneously at the {\it Z} Pulsed Power Facility \citep[e.g.,][]{McDaniel02,Matzen05,Rose10,Savage11} at Sandia National Laboratories.  ZAPP experiments take place in a large ($>60\,$m$^3$) vacuum chamber, each making use of the same {\it z}-pinch dynamic hohlraum \citep[e.g.,][]{Spielman98,Nash99,Stygar01,Sanford02,Bailey06,Rochau08} x-ray source to initiate plasma formation.

\begin{figure}[!h]
\includegraphics[width=\columnwidth]{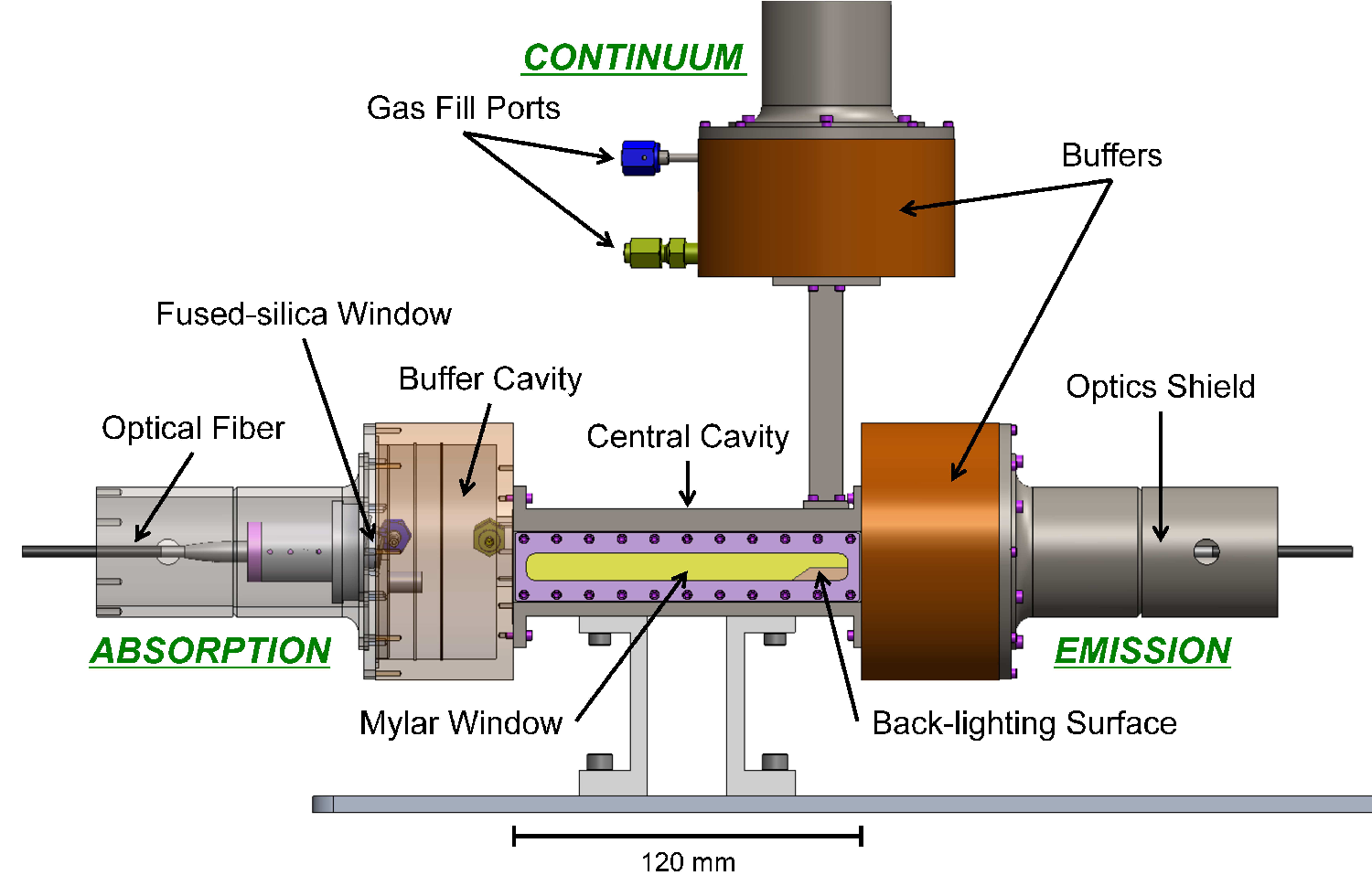}
\caption{False-color hardware drawing of our ``ACE''-configuration gas-cell assembly, capable of simultaneously observing (1) the H plasma in {\bf A}bsorption, (2) the {\bf C}ontinuum emission from a gold, back-lighting surface, and (3) the H plasma in {\bf E}mission.  For one line of sight we show the hardware within the buffer (orange) and optics shield (gray).}
\label{ACE}
\end{figure}

We place a gas-cell assembly (Figure~\ref{ACE}), filled with hydrogen (H$_2$), a distance away ($324\pm2\,$mm) from the x-ray source along a radial line of sight (LOS).  X-rays irradiate the cell, propagate through a thin ($1.44\pm0.02\,\mu$m) Mylar window and through the H$_2$ gas, and are absorbed by a gold wall at the back end of the cell cavity.  The gold heats to an electron temperature of $T_{\rm e}\sim$~few eV and re-emits as a continuum, which photoionizes the hydrogen.

\begin{figure}[!h]
\includegraphics[width=\columnwidth]{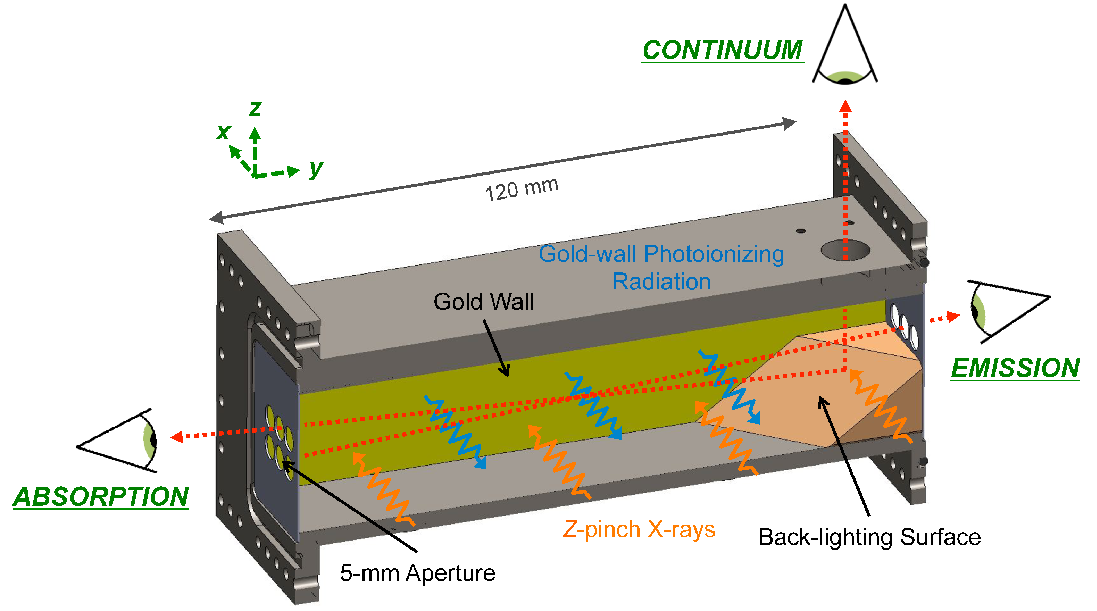}
\caption{Cross-section false-color hardware drawing of the gas-cell central cavity.  We illustrate the three observing lines of light (dashed, red lines).  The {\bf A}bsorption LOS observes the same area on the back-lighting surface as does the {\bf C}ontinuum LOS (vertical) but through the length of the plasma-filled cell.  The {\bf E}mission LOS also traverses this length but has no back-light.  Thus, both horizontal lines of sight observe the same nominal plasma self-emission.}
\label{central_cavity}
\end{figure}

We observe the H plasma along lines of sight parallel to the gold wall and perpendicular to the photoionizing radiation (Figure~\ref{central_cavity}).  For each LOS, a 50-mm-long buffer cavity separates the optics from the plasma formed within the central cavity.  Optical fibers deliver the light to time-resolved spectrometer systems.  Sections~\ref{transverse_gradient} and \ref{LOS_gradient} elaborate on our observations (Figures~\ref{transverse_schem} and \ref{boundary_schem} illustrate).  See \citet{Falcon13} for a description of our experimental platform and plasma formation, and see \citet{Falcon15c} for details of our data processing and calibration.

\section{Time-resolved Spectroscopic Measurements}

Inside the gas-cell central cavity a polyhedral stainless steel block, coated with 5\,$\mu$m of gold, rests on one end.  Its surface is tilted with respect to the plane normal to the {\it z}-pinch x-rays ($y$-$z$) and pitched with respect to the horizontal ($x$-$y$).  This allows the x-rays, the optics in one horizontal LOS ({\bf A}bsorption), and the optics in the vertical LOS ({\bf C}ontinuum) to each have a direct view.  The x-rays heat this surface like they do the gold back wall, allowing it to serve as a back-light for absorption measurements.

\subsection{Extracting Transmission}\label{extract}

Absorption spectra of plasmas contain three components:
\begin{equation}\label{eq_components}
I_\lambda^{\rm meas} = I_\lambda^{\rm back-light}T_\lambda^{\rm plasma} + I_\lambda^{\rm plasma}.
\end{equation}
$I_\lambda^{\rm back-light}$ is the spectral radiance of the back-lighting continuum.  The transmission, $T_\lambda^{\rm plasma}$, and the self-emission, $I_{\lambda}^{\rm plasma}$, both describe the plasma but in different ways.  We focus on $T_\lambda^{\rm plasma}$ for our analysis because of its preferable signal-to-noise (S/N) ratio.  It also provides a constraint on relative line strengths because all the transitions of the absorption lines we observe share the same initial (lower) state ($n=2$); this is a subject of future work.

The components in Equation~\ref{eq_components} are separately measurable using the ``ACE'' gas cell.  Because our absorption LOS ($I_\lambda^{\rm abs}$) observes a back-lighting surface through a length of plasma, $I_\lambda^{\rm abs}$ equals $I_\lambda^{\rm meas}$.  Our continuum LOS, which observes this same back-lighting surface (Figure~\ref{central_cavity}), also measures absorption spectra.  However, since the length of intervening plasma is minimal ($L\sim7\,$mm), the transmission is close to unity (absorption is small) and we assume negligible self-emission, so that $I_\lambda^{\rm cont}\approx I_\lambda^{\rm back-light}$.

With our emission LOS we approximate $I_\lambda^{\rm em}$ to equal $I_\lambda^{\rm plasma}$.  The length of plasma this LOS observes ($L\sim120\,$mm) nearly matches that of the absorption LOS ($L\sim114\,$mm), which is $\sim5\,$\% shorter.  We also approximate the nature of the plasma emission from each LOS to be the same, though the absorption LOS includes a $\sim$~few-mm region adjacent to a gold, back-lighting surface that the emission LOS does not (Figure~\ref{central_cavity}).

\begin{figure}[!h]
\includegraphics[width=\columnwidth]{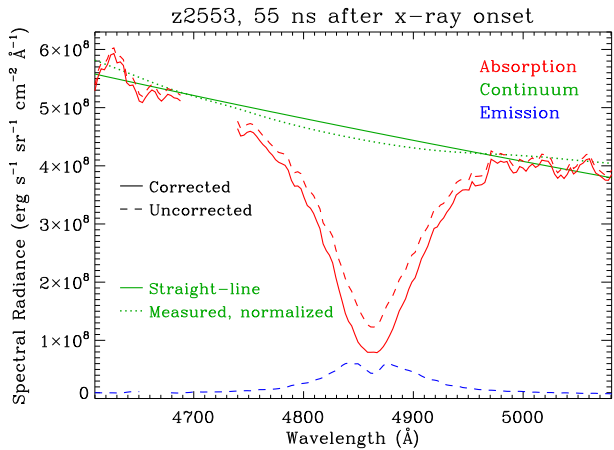}
\caption{Example H$\beta$ corrected-absorption spectrum (solid, red) plotted with its uncorrected-absorption (dashed, red) and with the subtracted-emission (dashed, blue).  These 10-ns line-outs center at 55\,ns after the onset of x-rays from experiment z2553.  We also show the back-lighting continuum shape measured from z2554 and normalized to the z2553 corrected-absorption.  This continuum shape (dotted, green) is similar to the straight-line continuum (solid, green) we use to convert these data to transmission.}
\label{emission_subtract}
\end{figure}

Upon placing our measured emission and absorption data onto the same absolute scale \citep{Falcon15c}, we correct the latter for plasma self-emission.  We can then substitute our measurements into Equation~\ref{eq_components} and invert to extract transmission:
\begin{equation}\label{eq_abscor}
T_\lambda^{\rm plasma}\approx \frac{I_\lambda^{\rm abs,cor}}{I_\lambda^{\rm cont}} \approx \frac{I_\lambda^{\rm abs} - I_\lambda^{\rm em}}{I_\lambda^{\rm cont}}.
\end{equation}
Figure~\ref{emission_subtract} shows an example H$\beta$ corrected-absorption spectrum (solid, red; $I_\lambda^{\rm abs,cor}$) versus its uncorrected spectrum (dotted, red; $I_\lambda^{\rm abs}$) for an integration of the time-resolved data over a 10-ns interval (line-out).

A desirable strategy to address plasma self-emission for absorption data measured for opacity studies is to require a back-lighter that is sufficiently bright to overwhelm self-emission \citep[e.g.,][]{Davidson88,Perry96,Bailey07,Bailey09,Bailey15}.  Our simultaneous emission and absorption measurements shed this requirement by allowing us to remove the former from the latter, and since we are the first to study H line profiles in absorption, we are also the first to utilize this technique for H.  Consequently, this strategy, and in particular the approximations we list, deserve further scrutiny.


At the electron densities we encounter, overlap of the H$\beta$ wings with adjacent lines (i.e., H$\alpha$ and H$\gamma$) is minimal.  Also, we do not observe the bound-free continuum \citep[e.g.,][]{Dappen87} to reach far enough redward to affect H$\beta$.  To focus on the H$\beta$ bound-bound transition we simplify our transmission extraction by redefining the continuum baseline ($I_\lambda^{\rm cont^*}$).  Since our measured shapes of the back-lighting continua (using the continuum LOS\footnotemark[3]) are quite linear in the H$\beta$ spectral vicinity (dotted, green in Figure~\ref{emission_subtract}), we use straight-line fits (solid, green); $I_\lambda^{\rm cont^*}\approx I_\lambda^{\rm cont}$.  This circumvents potential contamination due to any additional continuous (weakly dependent on wavelength) opacity sources, such as that due to the H$^-$ ion \citep[e.g.,][]{Wishart79,Griem97}.

\footnotetext[3]{Because only two spectrometer systems are sufficiently calibrated \citep{Falcon15c}, we do not use the continuum LOS to measure $I_\lambda^{\rm cont}$ ($I_\lambda^{\rm back-light}$) on experiment z2553.  This experiment uses one spectrometer system for the absorption LOS and the second for the emission LOS, enabling the self-emission correction.  On the subsequent experiment, z2554, the second spectrometer system uses the continuum LOS to measure the back-lighting spectral shape needed to infer transmission.}

\subsection{Spectral-Line Fitting Model}\label{fitting_pro}

Our fitting model is the transmission due to bound-bound absorption of a hydrogen Balmer line for a homogeneous (single $T_{\rm e}$, $n_{\rm e}$, and $n_{\rm l}$) plasma:
\begin{equation}
T_\lambda^{\rm b-b} = \psi * {\rm exp} \left\{ - \kappa_\lambda^{\rm b-b} L \right\},
\label{trans_eq}
\end{equation}
where $\psi$ is the measured instrumental broadening, $*$ is the convolution operator, $\kappa_\lambda^{\rm b-b}$ is the opacity due to photoexcitation for a bound-bound transition while neglecting stimulated emission \citep[e.g.,][]{Mihalas78book}, and $L$ is the length of the observed plasma.

We express the opacity as
\begin{equation}
\kappa_\lambda^{\rm b-b} = \frac{\pi e^2}{m_{\rm e} c}  f_{{\rm l}\to{\rm u}} n_{\rm l} w_{\rm u}(n_{\rm e})\phi_\lambda(T_{\rm e},n_{\rm e}),
\label{kappa_eq}
\end{equation}
where $e$ is the electron charge, $m_{\rm e}$ is the electron mass, $c$ is the speed of light, $f_{{\rm l}\to{\rm u}}$ is the oscillator strength of the transition from the lower to upper energy level, and $n_{\rm l}$ is the lower-level population.  The upper-level occupation probability, $w_{\rm u}(n_{\rm e})$, depends on electron density \citep{Hummer88}.  The line profile, $\phi_\lambda(T_{\rm e},n_{\rm e})$, is normalized ($\int \phi_\lambda{\rm  d}\lambda=1$) and a function of both electron temperature, $T_{\rm e}$, and $n_{\rm e}$.  Because the dependence on $T_{\rm e}$ is relatively weak (Section~\ref{electron_temperature}) and because we have empirical constraints (Section~\ref{measure_electron_temperature}), our fits assume $T_{\rm e}=1.0\,$eV.

Combining the constants into a single factor, $C_\kappa$, and specifying the H$\beta$ transition, Equation~\ref{trans_eq} becomes
\begin{equation}\label{H4equation_fit}
T_\lambda^{\rm H\beta} = \psi * {\rm exp} \left\{- C_\kappa  f_{2 \to 4} n_2 w_4 (n_{\rm e})\phi_\lambda^{\rm H\beta}(n_{\rm e}) L \right\}.
\end{equation}
The principal quantum numbers of the initial and final states are $n=2$ and $n=4$, respectively.

We use oscillator strengths from \citet{Baker08}, who reports to high precision ($<0.3\,\%$ uncertainty).  The occupation probability comes from \citet{Seaton90}; for H$\beta$, $w_{\rm u}(n_{\rm e})$ is quite nearly unity.  We fix $L$ according to the dimensions of our gas-cell hardware design.  This leaves $n_2$ and $\phi_\lambda^{\rm H\beta}(n_{\rm e})$ to be determined from experiment.

\subsection{Fits to Measured H$\beta$ Transmission}\label{fits}

\begin{figure*}[!b]
\centering
\includegraphics[width=\textwidth]{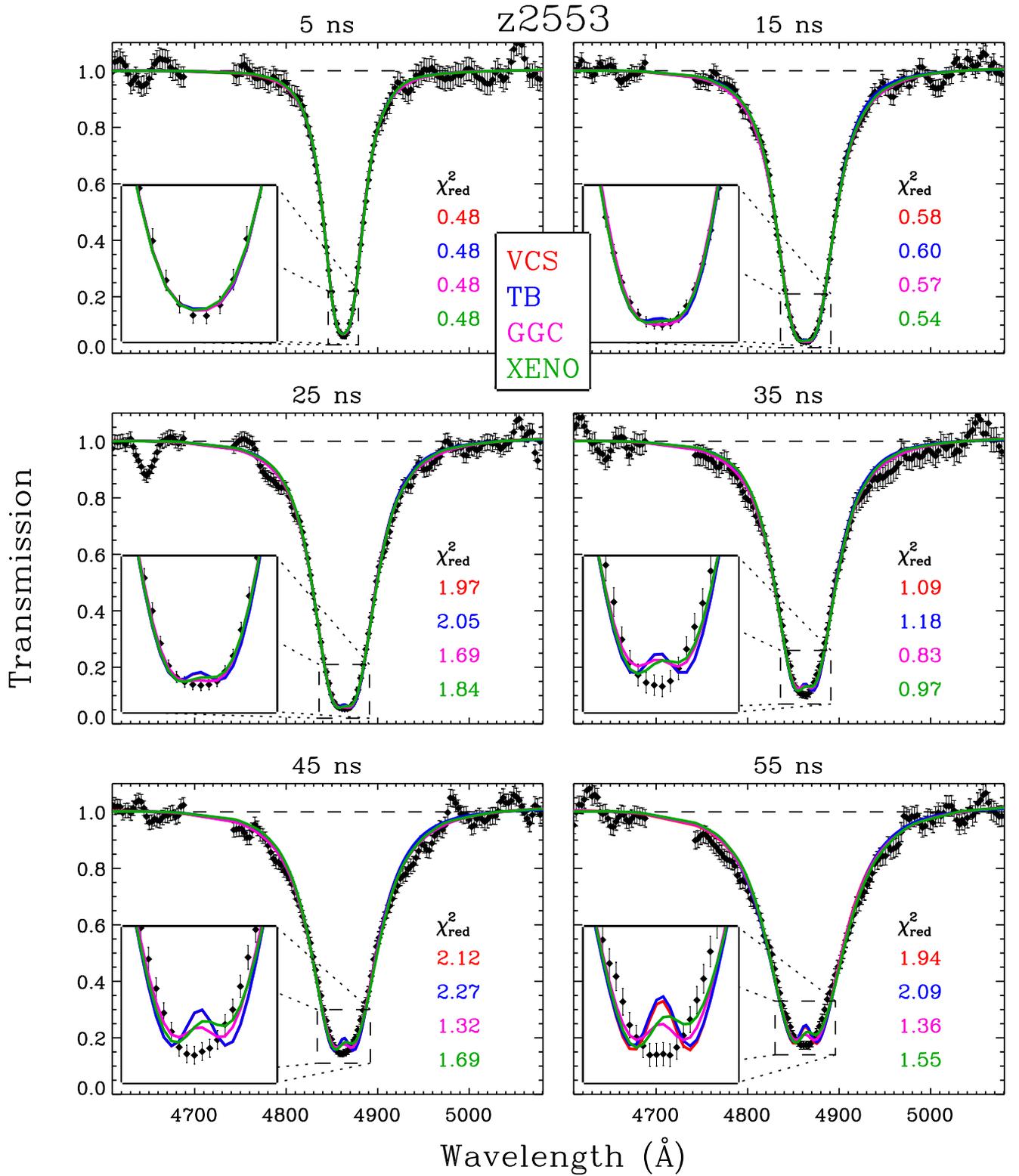}
\caption{Time sequence of H$\beta$ transmission spectra (black diamonds) measured from our laboratory plasma during experiment z2553.  Uncertainties (black, vertical lines) reflect the signal-to-noise ratio.  The first of these 10-ns line-outs begins at the onset of x-rays to the gas cell.  Fits (solid curves) use the theoretical line profiles of \citet[][VCS; red]{Vidal73}, \citet[][TB; blue]{Tremblay09}, \citet[][GGC; pink]{Gigosos03}, and \citet[][XENO; green]{Gomez15}.  We look closely at the line center where the fits differ most.}
\label{SixPlots_1}
\end{figure*}

\begin{figure*}[!t]
\centering
\includegraphics[width=\textwidth]{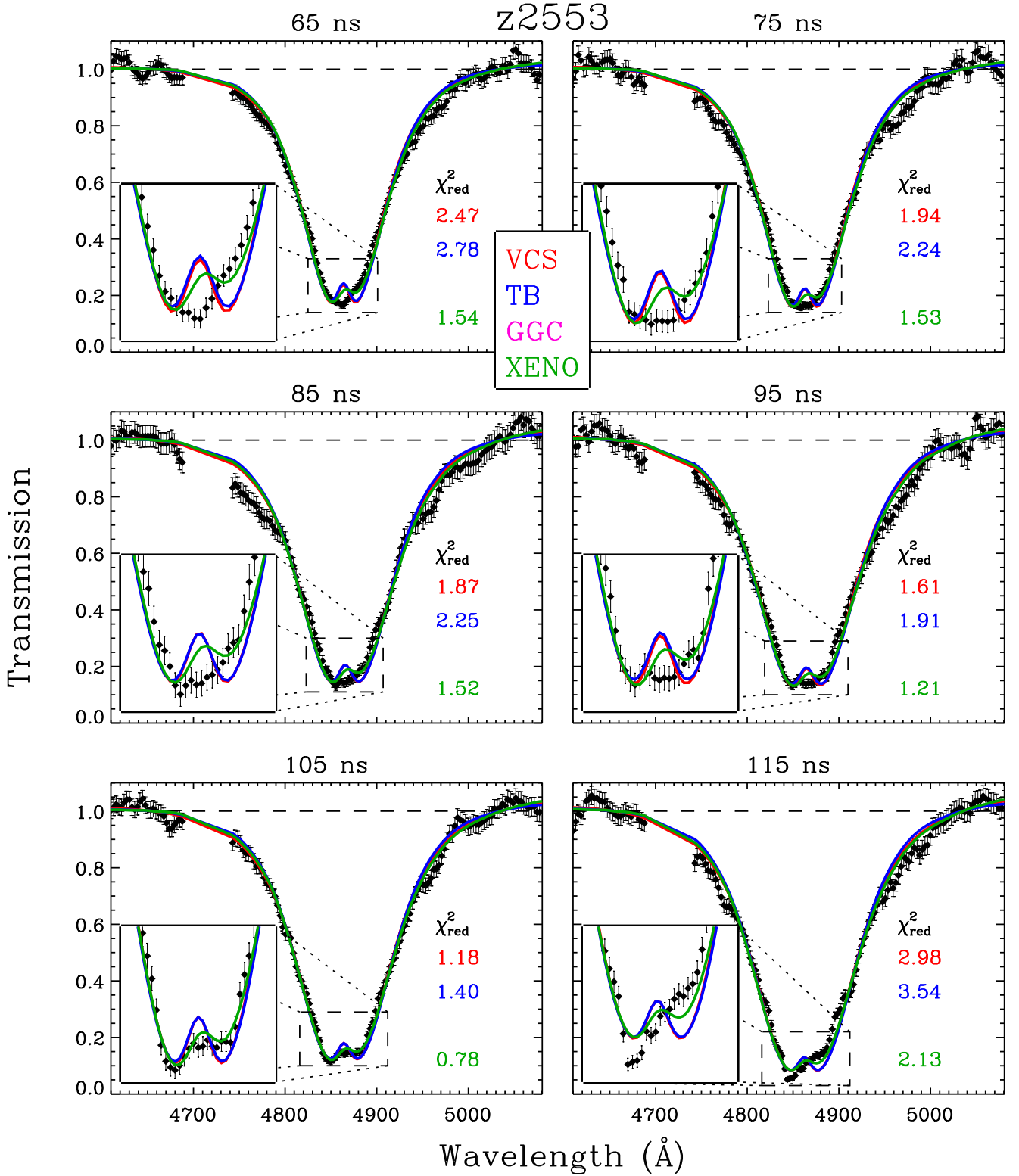}
\caption{Same as Figure~\ref{SixPlots_1} but plotting the subsequent six 10-ns line-outs.  Here we do not include GGC fits since the inferred electron densities exceed the range spanned by the GGC line-profile grid.}
\label{SixPlots_2}
\end{figure*}

Figures~\ref{SixPlots_1} and \ref{SixPlots_2} plot twelve consecutive 10-ns line-outs of H$\beta$ transmission (black diamonds) we measure from experiment z2553; the first line-out begins at the onset of x-rays (centered at 5\,ns after onset).  Uncertainties in spectral points (black, vertical lines) reflect the S/N.

We use a Levenberg-Marquardt minimization \citep{Levenberg44,Marquardt63} to fit our measured transmission spectra with Equation~\ref{H4equation_fit}.  Red, blue, pink, and green curves correspond to fits using theoretical line profiles, $\phi_\lambda^{\rm H\beta}$, that follow \citet[][VCS]{Vidal73}, \citet[][TB]{Tremblay09}, \citet[][GGC]{Gigosos03}, and \citet[][Xenomorph or XENO]{Gomez15}, respectively.  The VCS profiles come from \citet{Lemke97}.

These theoretical line profiles are quite similar, differing noticeably only in the line center (enlarged window in lower left-hand side of each panel).  Fits using VCS and TB -- both based on the Unified Theory --  are nearly identical.  Fits using GGC and Xenomorph -- both incorporating computer simulations \citep[e.g.,][]{Stamm84,Stambulchik10} -- exhibit less-pronounced central humps\footnotemark[4], and this latter theory is the only one of the four that calculates asymmetry.

\footnotetext[4]{The central feature of a line profile is usually described as the central {\it dip} because it is usually observed in emission.  In our novel approach, however, we observe the profile in absorption.}

\begin{figure}[!h]
\includegraphics[width=\columnwidth]{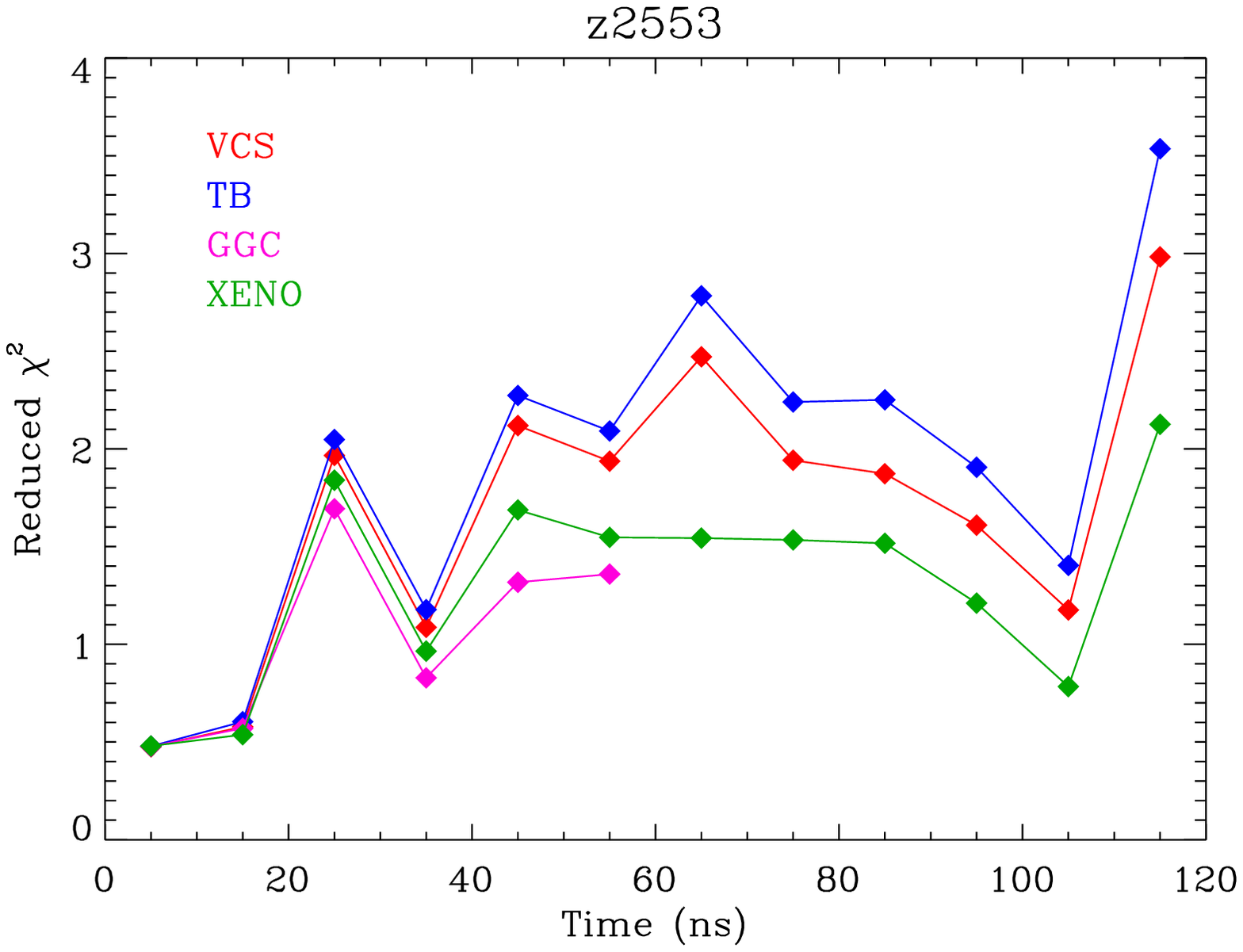}
\caption{Goodness-of-fit (reduced chi squared, $\chi_{\rm red}^2$) versus time for fits to H$\beta$ transmission measured from experiment z2553 and using VCS (red), TB (blue), GGC (pink), and Xenomorph (green) line profiles.  Early in time when $n_{\rm e}$ is low, all theories provide equally good fits.  Later in time when $n_{\rm e}\gtrsim10^{17}\,$cm$^{-3}$, GGC and Xenomorph profiles provide better fits than VCS and TB.}
\label{chi_vs_time}
\end{figure}

We indeed measure structure in the line center, including asymmetry \citep[e.g.,][]{Kudrin63,Wiese75,Halenka88,Djurovic05}.  This structure remains absent at the lowest $n_{\rm e}$ and increases with this parameter.  Early and late in the experiment the fits reproduce our measurements near the line center quite well, but from 35--95\,ns there are statistically significant deviations.  The fits predict central structure that we do not observe over this duration of the experiment, which spans the range of $1\times10^{17}\lesssim n_{\rm e}\lesssim3\times10^{17}\,$cm$^{-3}$.  Further work is required to ascertain whether this is due to deficiencies in the theoretical line profiles or in the experiment.

\begin{figure}[!h]
\includegraphics[width=\columnwidth]{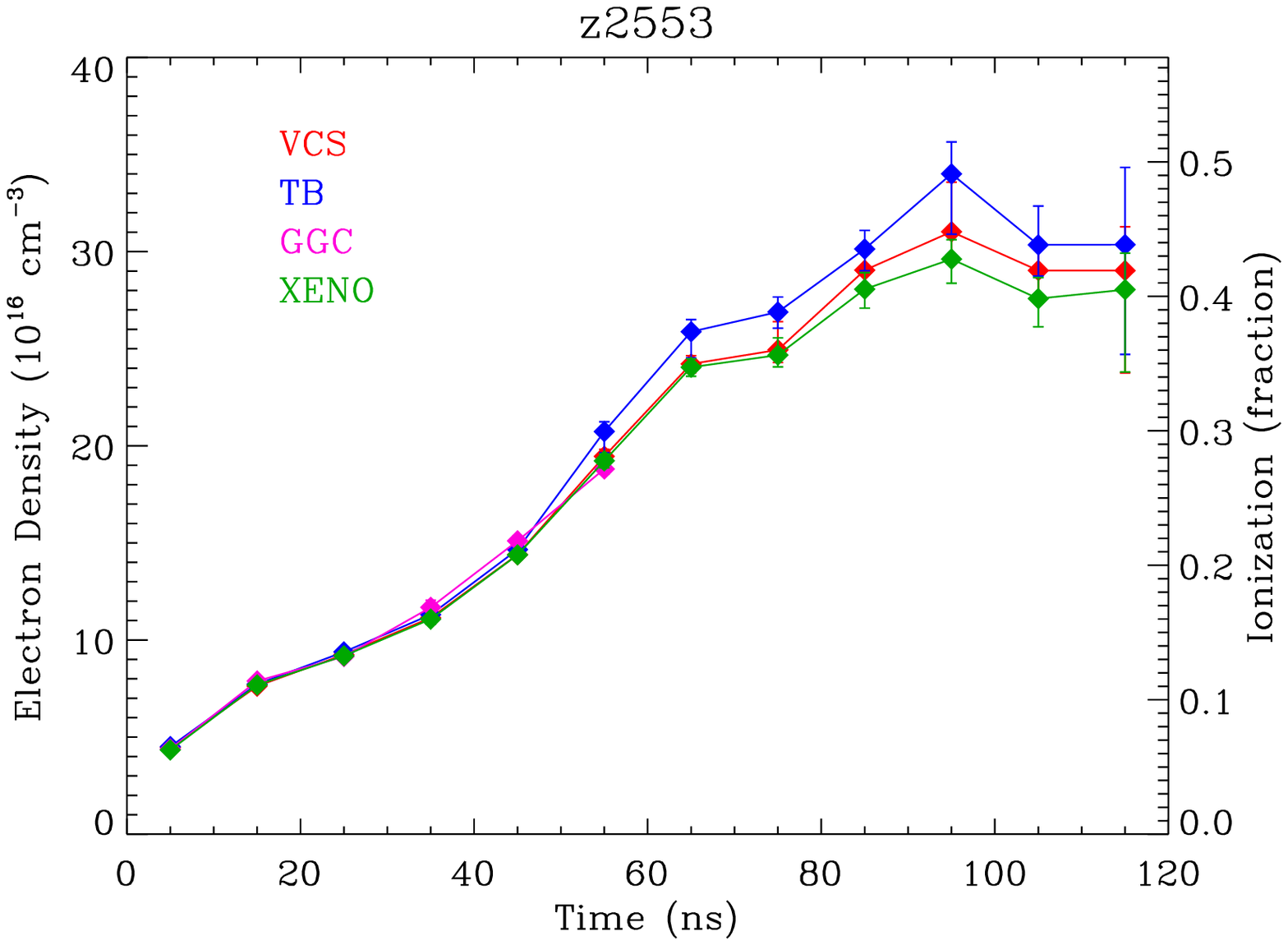}
\caption{Electron density (left-hand-side, vertical axis) and ionization (right-hand-side, vertical axis) versus time determined from fits to measured H$\beta$ transmission.  Section~\ref{fituncertainties} describes uncertainties (vertical lines).  We time-resolve a smooth increase in electron density ($n_{\rm e}$) which allows us to measure line profiles throughout a range of $n_{\rm e}$ in a single experiment and from the same plasma.  All theories infer $n_{\rm e}$ in agreement.}
\label{neH4_vs_time}
\end{figure}
\begin{figure}[!h]
\includegraphics[width=\columnwidth]{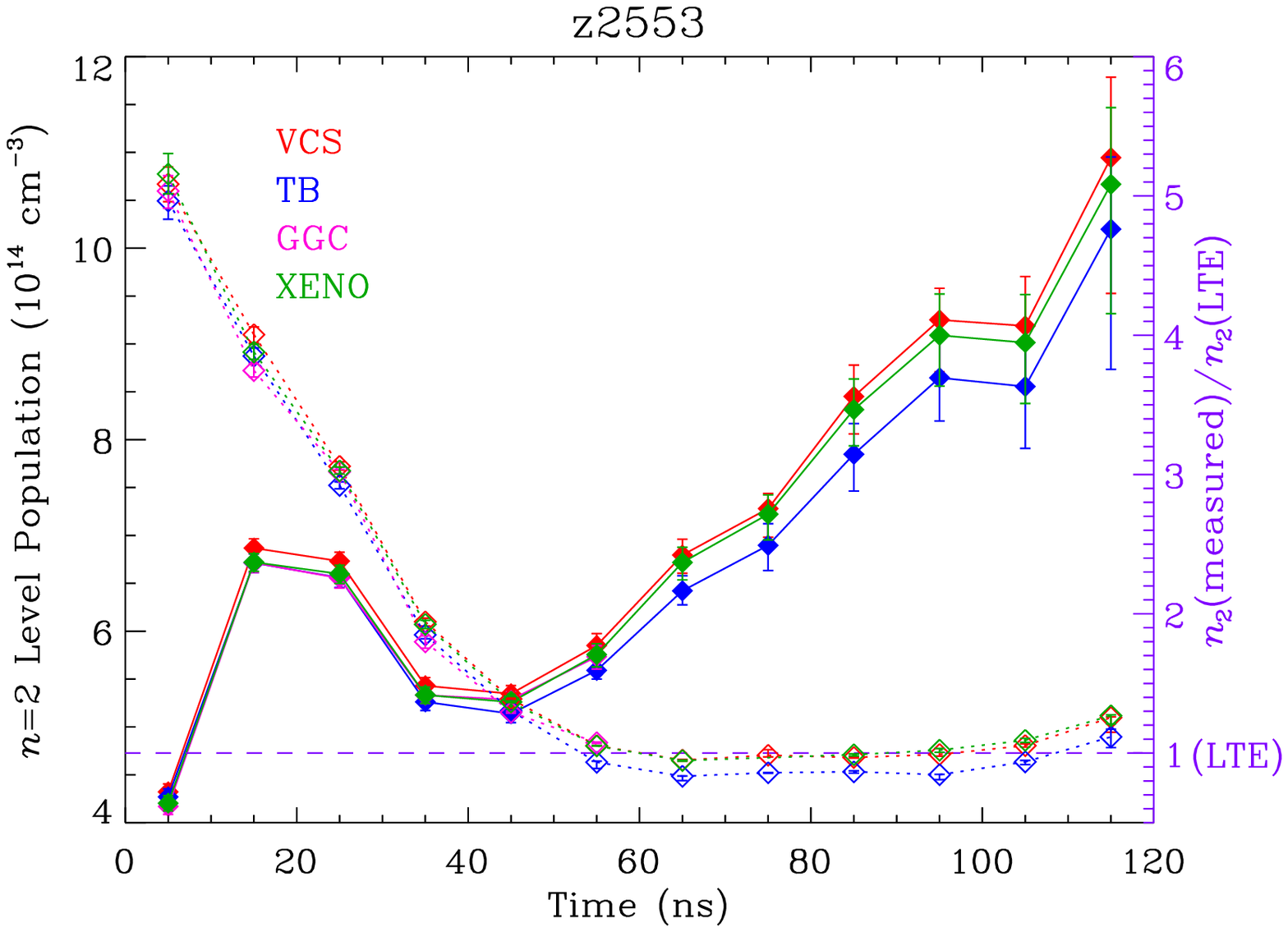}
\caption{Left-hand-side axis: similar to Figure~\ref{neH4_vs_time} but for lower ($n=2$) level population versus time.  Right-hand-side axis: ratio of measured versus LTE $n=2$ population (open diamonds, dotted lines).  By $\sim55\,$ns after the onset of x-rays the $n=2$ population is at its LTE value (dashed, purple, horizontal line).
\label{n2H4_vs_time}}
\end{figure}

GGC and Xenomorph profiles provide systematically better fits, as expressed by reduced $\chi_{\rm red}^2$ (Figure~\ref{chi_vs_time}).  Though the VCS and TB profiles yield poorer fits, all theories infer values for electron density, $n_{\rm e}$, and lower ($n=2$) level population, $n_2$, that agree (Figures~\ref{neH4_vs_time} and \ref{n2H4_vs_time}, respectively).  The standard deviation of $n_{\rm e}$ between theories increases from $\lesssim2$ to $\sim6\,$\% of $\langle n_{\rm e}\rangle$ as time increases (as $n_{\rm e}$ increases).  For $n_2$, the trend is the same, increasing from $\lesssim2$ to $\sim4\,$\%.

We monitor $n_{\rm e}$ increase throughout the first 95\,ns of experiment z2553 before reaching a plateau of $n_{\rm e}\sim3\times10^{17}\,$cm$^{-3}$.  Beyond 55\,ns we do not plot $n_{\rm e}$ inferred using GGC profiles, because these values extend beyond the $n_{\rm e}$ range covered by \citet{Gigosos03}.  In the first 25\,ns we overlap ($4\times10^{16}\lesssim n_{\rm e}\lesssim10^{17}\,$cm$^{-3}$) the electron-density range observed by \citet{Wiese72} -- the only other study to measure multiple H Balmer lines near these conditions.  We then exceed it by approximately a factor of three.

The right-hand-side, vertical axis of Figure~\ref{neH4_vs_time} shows the ionization, which we define as $\frac{n_{\rm e}}{n_{\rm e} + n^0}$, where $n^0$ is the neutral H atom density.  To determine the total particle density ($n_{\rm tot} = n_{\rm e} + n^0$), we measure the initial H$_2$-gas fill pressure inside our cell {\it in situ} using piezoresistive pressure sensors \citep[see Appendix A in][]{FalconPhDT}.  Assuming that the photoionizing radiation forming our plasma completely dissociates the H$_2$, the Ideal Gas Law translates our measured pressure ($P=10.63\pm0.05\,$Torr) to $n_{\rm tot}=(6.92\pm0.08)\times10^{17}\,$cm$^{-3}$.  Uncertainties for $P$ and $n_{\rm tot}$ reflect the precision of the pressure-sensor voltage measurement.  They do not include all systematic uncertainties, such as that due to the accuracy of the correction for zero-pressure output voltage.

\subsection{Electron-Temperature Determination}\label{measure_electron_temperature}

The lower-level population ($n_2$) also increases with time after exhibiting a moderate spike soon ($\sim20\,$ns) after the onset of x-rays (Figure~\ref{n2H4_vs_time}).  To give our measured $n_2$ context, we compare it to that of a plasma in local thermodynamic equilibrium (LTE) by plotting the ratio (open diamonds connected by dotted lines and corresponding to the purple, right-hand axis).  We determine the LTE $n=2$ population by first inserting our inferred electron density and measured total particle density into the Saha equation and solving for electron temperature.  We neglect ionization potential depression \citep[e.g.,][]{Crowley14} and approximate all neutral atoms to be in the ground state ($n^0\approx n_1$):
\begin{equation}
\frac{n_{\rm e}^2}{n_{\rm tot}-n_{\rm e}}\approx \frac{n_{\rm e}^2}{n_1} \approx \left( \frac{2\pi m_{\rm e}kT_{\rm e}}{h^2} \right)^{3/2}{\rm exp} \left\{ \frac{- \epsilon^0}{kT_{\rm e}}\right\},
\end{equation}
where $k$ is the Boltzmann constant, $h$ is the Planck constant, and the neutral-hydrogen ionization energy is $\epsilon^0=13.6\,$eV.
\begin{figure}[!h]
\includegraphics[width=\columnwidth]{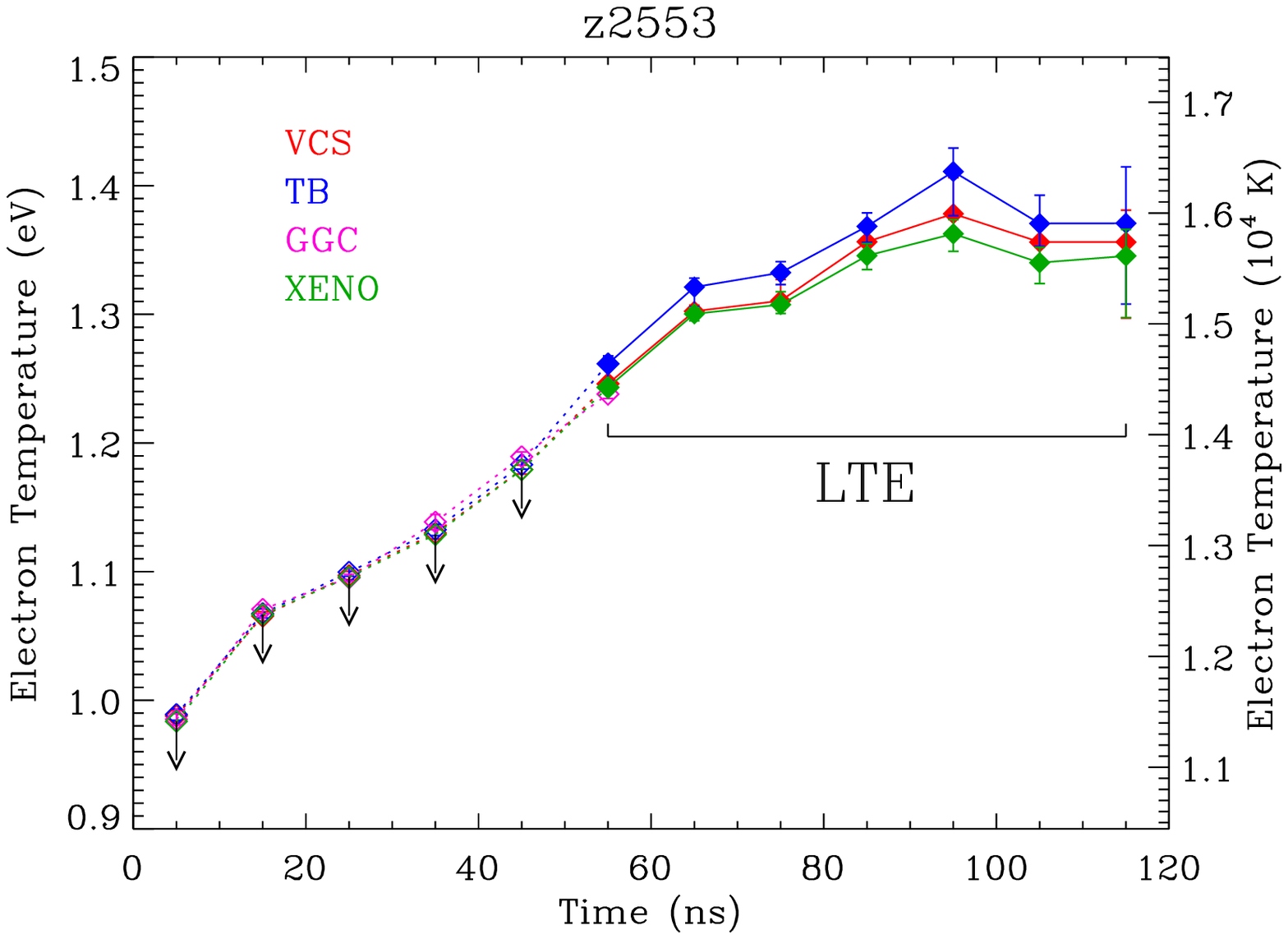}
\caption{Electron temperature ($T_{\rm e}$) versus time inferred from the Saha equation assuming local thermodynamic equilibrium (LTE) and using our measured ionization fraction (diamonds).  Before the plasma reaches LTE at $\sim55\,$ns (Figure~\ref{n2H4_vs_time}), these inferred $T_{\rm e}$ values are upper limits assuming the plasma is ``overionized'' (see text).
\label{T_vs_time}}
\end{figure}
Then, using this $T_{\rm e}$ with the Boltzmann relation yields $n_2({\rm LTE})$:
\begin{equation}
\frac{n_2({\rm LTE})}{n_1} = \frac{g_2}{g_1}{\rm exp} \left\{ \frac{-(\epsilon_2-\epsilon_1)}{kT_{\rm e}} \right\},
\end{equation}
where the statistical weight, $g_n$, equals $2n^2$, and the energy difference between the $n=2$ and $n=1$ states is $\epsilon_2-\epsilon_1=10.2\,$eV.  Early in time, our measured $n_2$ is $\sim5$ times that at LTE (dashed, purple, horizontal line).  This ratio ($\frac{n_2({\rm measured})}{n_2({\rm LTE})}$) monotonically approaches unity and reaches it $\sim55\,$ns after the onset of x-rays.  This implies that the ionization fraction, $n=2$ population, and $n=1$ population are consistent with their LTE values.

Once in LTE, $T_{\rm e}\sim1.25\,$eV ($\sim14,500\,$K) and rises to $T_{\rm e}\sim1.35\,$eV ($\sim15,700\,$K) by $\sim115\,$ns (Figure~\ref{T_vs_time}).  Before $\sim55\,$ns, our LTE $T_{\rm e}$ estimate provides an upper limit (open diamonds, dotted lines) to the true $T_{\rm e}$, assuming our plasma is ``overionized'' \citep[e.g.,][]{Kawasaki02}.  Our measurement of ``overexcitation'' (i.e., elevated $n=2$ population) supports this assumption, as does the premise that our plasma is initially photoionized \citep{Falcon13}.


\subsection{Uncertainties in Inferred Plasma Conditions}\label{fituncertainties}

The uncertainties for our inferred $n_{\rm e}$ (Figure~\ref{neH4_vs_time}), $n_2$ (Figure~\ref{n2H4_vs_time}), and, subsequently, $T_{\rm e}$ (Figure~\ref{T_vs_time}) have two sources, which we add in quadrature.  $\sigma^{\rm fit}$ is from the fit, which is a random uncertainty due to noise in the spectra.  It is typically $\sim2\,$\% and $\sim1\,$\% for $n_{\rm e}$ and $n_2$, respectively.  $\sigma^{\rm cal}$ is due to the transmission extraction (Section~\ref{extract}), which depends upon the accuracy of our calibration \citep[i.e., the relative measurement of plasma emission and absorption from different spectrometer systems;][]{Falcon15c}.

Early in time, the magnitude of $\sigma^{\rm cal}$ is similar to that of $\sigma^{\rm fit}$.  Throughout the experiment, however, $\sigma^{\rm cal}$ increases to $\sim$~several \%.  This is a result of the evolution of the back-lighting continuum emission ($I_\lambda^{\rm cont}$) which decreases as the gold surface cools, thus increasing the significance of the self-emission component ($I_\lambda^{\rm em}$) to the measured absorption ($I_\lambda^{\rm abs}$).


\subsection{Plasma Reproducibility}

\begin{figure}[!h]
\includegraphics[width=\columnwidth]{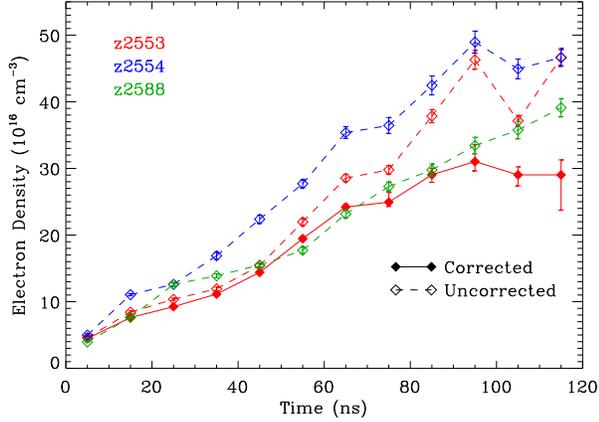}
\caption{Similar to Figure~\ref{neH4_vs_time}.  The inferred electron density versus time from multiple experiments is qualitatively reproducible.  Also, fits to H$\beta$ transmission uncorrected (open diamonds, dashed lines) for plasma self-emission overestimate the electron density compared to fits with the correction (filled diamonds, solid lines).  We show fits using VCS theoretical line profiles.
\label{neH4_vs_time_reprod}}
\end{figure}

We observe the same qualitative trends in goodness-of-fit, electron density (Figure~\ref{neH4_vs_time_reprod}), and lower-level population versus time across multiple experiments, giving credence to plasma reproducibility using our experimental platform.  This includes the $n_2$ spike at $\sim20\,$ns.

Experiments z2554 (blue) and z2588 (green) have no $I_\lambda^{\rm em}$ measurement, so we infer $n_{\rm e}$ from fits to $I_\lambda^{\rm abs}$ (i.e., transmission uncorrected for plasma self-emission; open diamonds, dashed lines).  For z2553 (red), foregoing the emission correction overestimates $n_{\rm e}$ by $\sim$~few\,\% early in time, increasing to $\sim$~few tens of \% late in time.  It also underestimates $n_2$ by percentages similar to the $n_{\rm e}$ overestimations.

\section{Synthetic Investigations into Systematic Uncertainties}

Our fitting procedure (Section~\ref{fitting_pro}) approximates the line profile, $\phi_\lambda^{\rm H\beta}$, to have a negligible dependence on electron temperature, $T_{\rm e}$, and therefore depend solely on electron density, $n_{\rm e}$.  It also assumes that our experimental plasma either (1) is homogeneous and exists at a single plasma condition or (2) can be sufficiently and uniquely described as a homogeneous plasma.

We implement synthetic data analysis \citep[e.g.,][]{Nagayama12,Nagayama12b} to investigate the sensitivity of our fits to the aforementioned factors.  We simulate H$\beta$ transmission data at different $T_{\rm e}$ (Section~\ref{electron_temperature}) as well as those resulting from inhomogeneous conditions (Sections~\ref{transverse_gradient} and \ref{LOS_gradient}).  These data use VCS theoretical line profiles, and include random noise (S/N $=33$) and instrumental broadening (Voigt profile with $\sim10$-\AA\ full-width-at-half-maximum) typical for our experiments.  They also span a range of plasma conditions (i.e., $10^{16}\le n_{\rm e}\le 4\times10^{17}\,$cm$^{-3}$ in steps of $\Delta n_{\rm e}=3\times10^{16}\,$cm$^{-3}$ and $3\times10^{14}\le n_2\le 1.2\times10^{15}\,$cm$^{-3}$ in steps of $\Delta n_2=10^{14}$ or $3\times10^{14}\,$cm$^{-3}$).  We independently adjust $T_{\rm e}$, $n_{\rm e}$, and $n_2$ without assuming LTE.


\subsection{Sensitivity to Electron Temperature}\label{electron_temperature}

\begin{figure}[!h]
\includegraphics[width=\columnwidth]{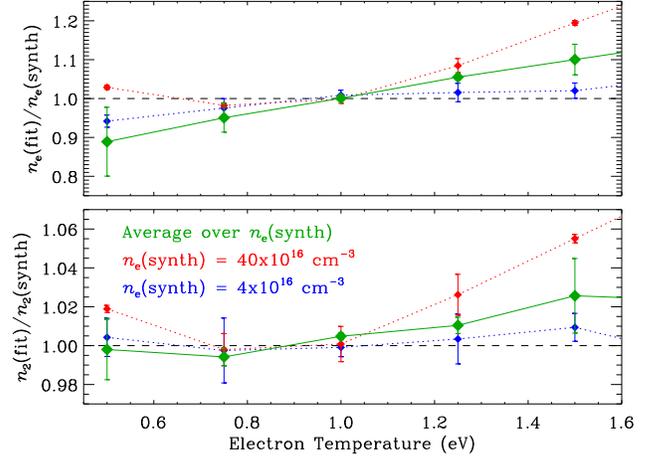}
\caption{Ratio (vertical axis) of inferred parameter (fit) and the value of that parameter for synthetic H$\beta$ transmission (synth) created at different electron temperatures (horizontal axis).  Our fits assume $T_{\rm e}=1.0\,$eV.  Top and bottom panels show inferred electron density, $n_{\rm e}$, and lower ($n=2$) level population, $n_2$, respectively.}
\label{T_fig}
\end{figure}

Our ionization fraction ($\frac{n_{\rm e}}{n_{\rm tot}}$) and $n_2$ measurements indicate that our plasma reaches LTE by $\sim55\,$ns (Figure~\ref{n2H4_vs_time}).  This corresponds to a thermalized temperature of $T_{\rm e}\sim1.3\,$eV.  Before this time and while our plasma is not in LTE, we assume that our $T_{\rm e}$ estimates are upper limits (Figure~\ref{T_vs_time}); a precise $T_{\rm e}$ determination is nontrivial, likely requires collisional-radiative hydrodynamic modeling \citep[e.g.,][]{HansenS07}, and is beyond the scope of this paper.  Because of this and because of the relative insensitivity we show in Figure~\ref{T_fig}, we adopt a compromise (as a function of time and hence $T_{\rm e}$) by fixing $T_{\rm e}=1.0\,$eV in our spectral-line fitting model (Section~\ref{fitting_pro}).

By assuming $T_{\rm e}=1.0\,$eV, our fits underestimate $n_{\rm e}$ (top panel) and $n_2$ (bottom panel) for cooler ($T_{\rm e}<1.0\,$eV) plasmas  and overestimate for warmer ($T_{\rm e}>1.0\,$eV) plasmas.  This effect is more significant for $n_{\rm e}$ than for $n_2$, for which it is quite small ($\lesssim1\,$\% for most cases).   Uncertainties (vertical lines) only simulate $\sigma^{\rm fit}$ and do not include $\sigma^{\rm cal}$ (Section~\ref{fituncertainties}).

\subsection{Gradient in Electron Density Transverse to our Line of Sight}\label{transverse_gradient}

As a consequence of the plasma formation inside our gas cell, ionization decreases with increasing distance from the gold wall \citep{Falcon13}.  Since we observe our plasma parallel to the gold wall (perpendicular to the photoionizing radiation) along lines of sight with finite diameter \citep[$\sim3\,$mm;][]{Falcon15c}, our measured spectra sample some range of plasma conditions.

\begin{figure}[!h]
\includegraphics[width=\columnwidth]{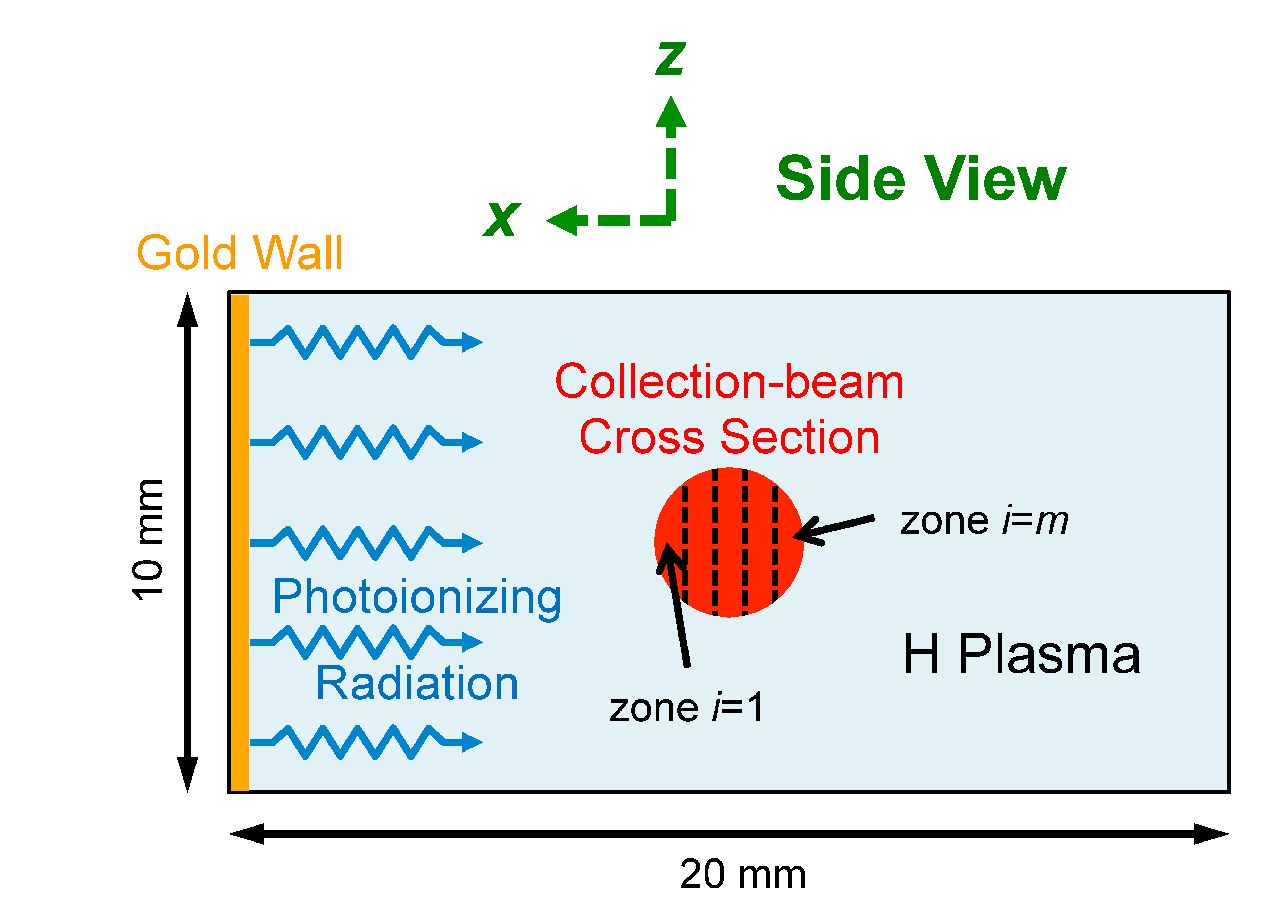}
\caption{Cross-section schematic of the gas-cell central cavity from the point of view of a horizontal line of sight.  The collection beam (red) traverses the hydrogen plasma parallel to the gold wall and perpendicular to the photoionizing radiation.  We construct synthetic data simulating a transverse gradient in electron density, $n_{\rm e}$, by assembling the collection beam out of zones with different $n_{\rm e}$ linearly decreasing with distance from the gold wall ($x$ direction).}
\label{transverse_schem}
\end{figure}

We simulate a linear gradient in $n_{\rm e}$ transverse ($x$ direction) to our LOS using the mean of synthetic homogeneous transmission spectra of different $n_{\rm e}$ weighted according to fractional areas of a circle.  This corresponds to our cylindrical collection beam.  We describe the gradient as the difference of $n_{\rm e}$ at the boundaries of the LOS collection beam divided by the mean $n_{\rm e}$:
\begin{equation}\label{trans_gradient_eq}
{\rm Gradient}\equiv\frac{|n_{\rm e}(i=m)-n_{\rm e}(i=1)|}{\frac{1}{m}\sum\limits_{i=1}^{m} n_{\rm e}(i)}\times100\,\%,
\end{equation}
where zone $i$ is one of $m$ zones (Figure~\ref{transverse_schem}).

\begin{figure}[!h]
\includegraphics[width=\columnwidth]{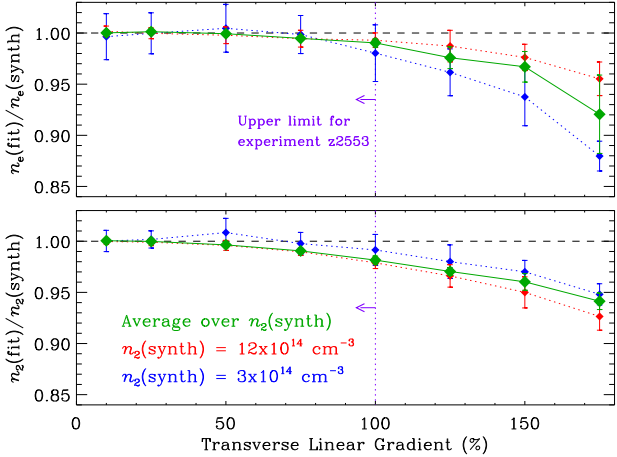}
\caption{Vertical axis: ratio of inferred parameter (fit) and the value of that parameter used to create synthetic H$\beta$ transmission (synth) that simulates a linear gradient in electron density, $n_{\rm e}$, transverse ($x$ direction) to the observed line of sight (LOS).  Horizontal axis: $n_{\rm e}$ gradient transverse to synthetic LOS defined in Equation~\ref{trans_gradient_eq}.  Top and bottom panels show inferred $n_{\rm e}$ and lower ($n=2$) level population, $n_2$, respectively.  The steeper the gradient, the more the fit underestimates each parameter.}
\label{transversegrad_fig}
\end{figure}

The top and bottom panels of Figure~\ref{transversegrad_fig} plot our results for $n_{\rm e}$ and $n_2$, respectively.  We show the ratio (colored diamonds) of the parameter inferred from fitting and the synthetic value the fit sought to recover (i.e., $\frac{n_{\rm e}({\rm fit})}{n_{\rm e}({\rm synth})}$ and $\frac{n_2({\rm fit})}{n_2({\rm synth})}$).  For $n_{\rm e}$, this synthetic value is the mean $n_{\rm e}$ across zones, and, since the gradient is linear, it equals that of the central zone.

As the gradient increases, fits underestimate both $n_{\rm e}$ and $n_2$.  Empirically, we find that neither $\frac{n_{\rm e}({\rm fit})}{n_{\rm e}({\rm synth})}$ nor $\frac{n_2({\rm fit})}{n_2({\rm synth})}$ depend on $n_{\rm e}$, so we average over this parameter.  They do depend on $n_2$, however.  As $n_2$ (line strength) increases, fits underestimate $n_{\rm e}$ less and $n_2$ more.

Though an exponential function better describes the decrease in measured $n_{\rm e}$ with increasing distance from the gold wall \citep{Falcon13}, we approximate this gradient to be linear across our $3$-mm-diameter collection beam.  The measured gradient from experiments z2300 and z2302 is then $\sim100\,$\% for a LOS centered at 10-mm away from the gold wall and using a $\sim30$-Torr H$_2$ gas fill \citep{Falcon13}.  Our synthetic test suggests that this could lead to an underestimate of both $n_{\rm e}$ and $n_2$ of up to $\sim3\,$\%.

Experiment z2553 uses a $\sim10$-Torr gas fill.  Lower pressure results in decreased attenuation of the photoionizing radiation with increasing distance from the gold wall.  This means we encounter higher $n_{\rm e}$ for a fixed LOS distance.  It also means the transverse gradient is not as steep.  Therefore, the measured $\sim100$-\% gradient should be an overestimate for the z2553 observing LOS (dotted, purple, vertical line).  The underestimation of $n_{\rm e}$ and $n_2$ is then $\lesssim3\,$\%.

\subsection{Gradient in Electron Density Along our Line of Sight or a Boundary-Layer Plasma}\label{LOS_gradient}

An inevitable attribute of laboratory plasmas is their finite size.  Boundary layers exist where conditions transition away from that of the bulk plasma, contributing to inhomogeneity along the LOS.  Observing long plasmas minimizes this effect \citep[e.g.,][]{Bengtson76}, and though we create the longest photoionized (radiation-driven) laboratory plasmas to date \citep{Falcon13}, their lengths remain finite.

\begin{figure}[!h]
\includegraphics[width=\columnwidth]{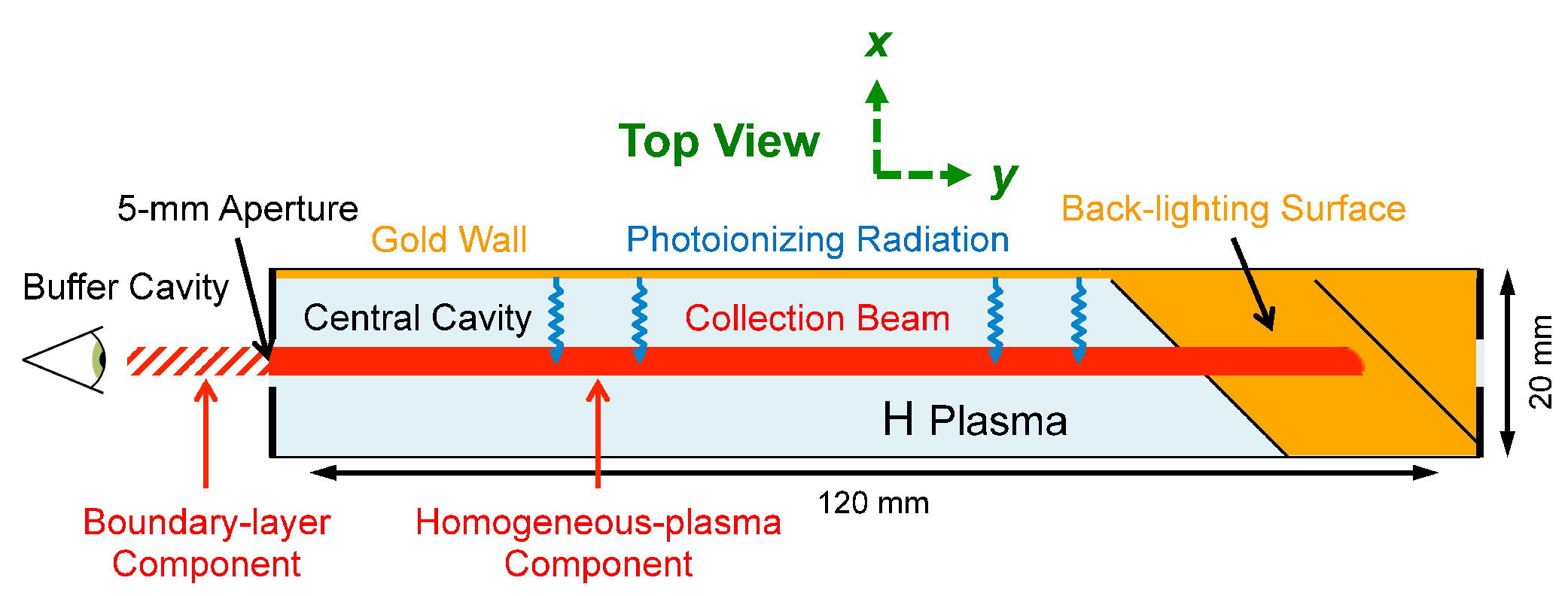}
\caption{Top-view cross-section schematic of the gas-cell central cavity.  We construct synthetic data simulating a line-of-sight gradient in electron density by appending a foreground boundary layer (diagonally striped, red) to the homogeneous-plasma component (solid, red) within a collection-beam volume.}
\label{boundary_schem}
\end{figure}

We assume that the boundary-layer plasma extending into the buffer cavity (Figure~\ref{ACE}) is the dominant source of inhomogeneity along our observing LOS  ($y$ direction).  Thus, we create synthetic data consisting of two components: a homogeneous plasma (solid, red in Figure~\ref{boundary_schem}) the length ($L\sim114\,$mm) of the absorption LOS inside our gas-cell central cavity, and a boundary layer (diagonally striped, red) with variable length whose $n_{\rm e}$ linearly decreases along that length toward the observer.  This homogeneous component neglects the region adjacent to the gold, back-lighting surface (Section~\ref{extract}), and the boundary-layer $n_{\rm e}$ does not decrease to zero from the electron density, $n_{\rm e}({\rm synth})$, of the homogeneous component.  It decreases to the lowest value of our theoretical line-profile grid, $n_{\rm e}=10^{15}\,$cm$^{-3}$.

\begin{figure}[!h]
\includegraphics[width=\columnwidth]{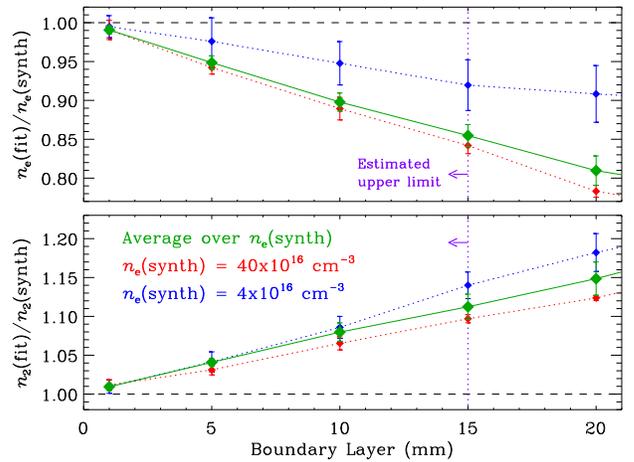}
\caption{Vertical axis: ratio of inferred parameter (fit) and the value of that parameter used to create synthetic H$\beta$ transmission (synth) that simulates a boundary-layer plasma in the foreground of a homogeneous plasma.  Horizontal axis: length of synthetic boundary layer.  The longer this layer, the more each fit underestimates $n_{\rm e}$ (top panel) and overestimates $n_2$ (bottom panel).}
\label{LOSgrad_fig}
\end{figure}

A synthetic boundary layer in the foreground of a homogeneous plasma causes fits to underestimate $n_{\rm e}$ and overestimate $n_2$ (Figure~\ref{LOSgrad_fig}).  Longer boundary layers exacerbate these effects.  Empirically, they do not depend on $n_2$ (similarly to the lack of $n_{\rm e}$-dependence of $\frac{n_{\rm e}({\rm fit})}{n_{\rm e}({\rm synth})}$ and $\frac{n_2({\rm fit})}{n_2({\rm synth})}$ for the transverse gradient in Section~\ref{transverse_gradient}), so we average over this parameter.  As $n_{\rm e}(\rm synth)$ increases, fits underestimate $n_{\rm e}$ more and overestimate $n_2$ less.

With current measurements, the true conditions in the buffer cavity are difficult to ascertain.  As a conservative estimate, though, we do not expect a boundary-layer plasma to extend beyond $\sim15\,$mm (dotted, purple, vertical line) into the 50-mm-long buffer.  In similar experiments we measure $n_{\rm e}$ to significantly fall off at 15\,mm from the gold wall \citep{Falcon13}.  This is in the $x$ direction, perpendicular to the gold wall, where the LOS to the photoionizing radiation is direct, thus maximizing the irradiance.  Not only is the LOS to a boundary layer peripheral, a 5-mm-diameter aperture partitions the central cavity from the buffer cavity (Figures~\ref{central_cavity} and \ref{boundary_schem}), choking the radiation that can penetrate.  At 15\,mm our synthetic test suggests that we underestimate $n_{\rm e}$ and overestimate $n_2$ by $\lesssim14\,$\%.

\section{Conclusions}

We examine hydrogen Balmer-$\beta$ spectral line profiles measured from laboratory plasmas at white dwarf photospheric conditions; we establish this line as a diagnostic anchor for our experiments. This sets the stage for the analysis of the relative line shapes and strengths of multiple H Balmer lines we measure simultaneously (H$\beta$, H$\gamma$, and H$\delta$).

The conditions we encounter during a single experiment smoothly increase throughout a 120-ns plasma evolution and span from $n_{\rm e}\sim4$ to $\sim30\times10^{16}\,$cm$^{-3}$, exceeding, by a factor of three, the highest $n_{\rm e}$ achieved in the benchmark work of \citet{Wiese72}.  At $n_{\rm e}\gtrsim10^{17}\,$cm$^{-3}$, we find that the theoretical line profiles adopted throughout the WD astronomy community \citep[i.e.,][]{Lemke97,Tremblay09} do not fit our measured H$\beta$ profiles as well as computer-simulation-based calculations \citep[i.e.,][]{Gigosos03,Gomez15}.  Despite this, all these profiles infer similar $n_{\rm e}$ and $n=2$ level populations, $n_2$; the standard deviations of inferred parameters between theories remain modest, increasing, as $n_{\rm e}$ increases, from $\lesssim2$ to $\sim6\,$\% of $\langle n_{\rm e}\rangle$ and from $\lesssim2$ to $\sim4\,$\% of $\langle n_2\rangle$.

We investigate the sensitivity of our spectral-line fitting model to the following factors: (1) the approximation that the H$\beta$ line shape has a negligible dependence on electron temperature, (2) a gradient in $n_{\rm e}$ transverse to our observing line of sight (LOS), and (3) an interloping boundary-layer plasma along our LOS.

For each of these we have experimental guidance.  (1): By monitoring $n_2$ we witness our photoionized plasma smoothly relax into local thermodynamic equilibrium (LTE) in $\sim55\,$ns, which corresponds to $T_{\rm e}\sim1.3\,$eV ($\sim15,000\,$K).  Since our fits assume $T_{\rm e}=1.0\,$eV, we may overestimate $n_{\rm e}$ by $\lesssim10\,$\% (at our highest $n_{\rm e}$) and $n_2$ by $\lesssim3\,$\%.  (2) and (3): Using this experimental platform, we measure $n_{\rm e}$ of a similar H plasma along lines of sight at different distances from the radiating gold wall \citep{Falcon13}.  This constrains the transverse gradient across our LOS for this experiment to be $\lesssim100\,$\%.  It also suggests that a boundary-layer plasma along our LOS is, conservatively, $\lesssim10\,$\% of our homogeneous-plasma length.   Due to (2), we should underestimate both $n_{\rm e}$ and $n_2$ by $\lesssim3\,$\%.  Due to (3) we should underestimate $n_{\rm e}$ and overestimate $n_2$ by $\lesssim14\,$\%.


Our ability to measure multiple H lines simultaneously at high electron densities ($n_{\rm e}>10^{17}\,$cm$^{-3}$) provides us with a sensitive and unprecedented test of theoretical line profiles.  Relative H Balmer line {\it shapes} can discriminate between theories \citep[e.g.,][]{Vidal73,Tremblay09}; because we measure in {\it absorption}, relative line {\it strengths} will offer a new perspective on occupation probabilities \citep{Hummer88}.

\acknowledgements 
This work was performed at Sandia National Laboratories and is supported by the Laboratory Directed Research and Development program.  We thank the {\it Z} Facility teams and in particular, D.~Bliss, A.~Carlson, T.~Lockard, L.~Nielsen-Weber, and D.~Sandoval, for experimental support.  Sandia is a multiprogram laboratory operated by Sandia Corporation, a Lockheed Martin Company, for the United States Department of Energy under contract DE-AC04-94AL85000.  We thank P.-E. Tremblay for providing TB theoretical line profiles.  We also thank R.~Bengtson, S.~Hansen, and G.~Loisel for useful discussions and A.~Wootton for championing our fundamental-science research effort.  R.E.F., M.H.M., and D.E.W. acknowledge support from the United States Department of Energy under grant DE-SC0010623.  T.A.G acknowledges support from the National Science Foundation Graduate Research Fellowship under grant DGE-1110007.  This work has made use of NASA's Astrophysics Data System Bibliographic Services.

\bibliographystyle{yahapj}
\bibliography{/Users/refalco/Desktop/Writing/all}

\end{document}